\documentclass[onecolumn]{revtex4}
\usepackage{amsmath}

\makeatletter
\def\btt#1{\texttt{\@backslashchar#1}}%
\DeclareRobustCommand\bblash{\btt{\@backslashchar}}%
\makeatother

\begin{document}

\title{Comment on 'Schwarzschild Black Hole in Noncommutative Spaces'}
\

\author{Xin-zhou Li}\email{kychz@shtu.edu.cn}

\affiliation{Shanghai United Center for Astrophysics(SUCA),\\
 Shanghai Normal University, 100 Guilin Road, Shanghai 200234,China
}%

\date{\today}

\begin{abstract}
\section*{abstract}
A brief comment on the paper( arXiv: hep-th/0508051)with the
mentioned title by F. Nasseri.
\end{abstract}


\maketitle

  Recently there has been considerable interest in the possible
  effects of the noncommutative space\cite{Seiberg,Hewett,Aschieri,Carroll,li,liu}. In the noncommutative
  quantum mechanics\cite{Chaichian}, one has assumed the commutation rules as
  follows
\begin{eqnarray}\label{commutation}
&&[\hat{x}^{i},\hat{x}^{j}]=i\theta_{ij}\;,\nonumber\\
&&[\hat{x}_{i},\hat{p}_{j}]=i\delta_{ij}\;,\\
&&[\hat{p}_{i},\hat{p}_{j}]=0\;\nonumber.
\end{eqnarray}
\noindent and the coordinate transformation
\begin{eqnarray}\label{transformation}
&&x_{i}=\hat{x}_{i}+\frac{1}{2}\theta_{ij}\hat{p}_{j}\;,\nonumber\\
&&p_{i}=\hat{p}_{i}\;.
\end{eqnarray}
\noindent where the new variables satisfy the usual canonical
commutation rules of quantum mechanics:
\begin{eqnarray}\label{quantum}
&&[x_{i},x_{j}]=0\;,\nonumber\\
&&[x_{i},p_{j}]=i\delta_{ij}\;,\\
&&[p_{i},p_{j}]=0\;.\nonumber
\end{eqnarray}

By using these commutation rules, the author of Ref.\cite{Nasseri}
argued that the Schwarzschild black hole can be extended to the
noncommutative space. Here, the basic concept were thrown into
confusion because the Schwarzschild external metric is a classical
solution which has nothing to do with the commutation rules of
quantum mechanics in the noncommutative or commutative space. In
fact, one has to develope the modified Poisson brackets of
classical phase space $(\tilde{x}_{i},\tilde{p}_{i})$ if one would
like to consider that classical metric extends to the
noncommutative case. Let us now infer the Poisson brackets from
the rules of commutation of noncommutatitive quantum mechanics
(\ref{commutation}),
\begin{eqnarray}\label{commutation2}
&&\{\tilde{x}_{i},\tilde{x}_{j}\}=\theta_{ij}\;,\nonumber\\
&&\{\tilde{x}_{i},\tilde{p}_{j}\}=\delta_{ij}\;,\\
&&\{\tilde{p}_{i},\tilde{p}_{j}\}=0\;\nonumber\;
\end{eqnarray}
and the modified Poisson brackets for two arbitrary functions $F$
and $G$ defined on the phase space
\begin{eqnarray}\label{function}
&&\{F,G\}=\theta_{ij}\frac{\partial F}{\partial
\tilde{x}_{i}}\frac{\partial G}{\partial
\tilde{x}_{j}}+(\frac{\partial F}{\partial
\tilde{x}_{i}}\frac{\partial G}{\partial
\tilde{p}\,^{i}}-\frac{\partial F}{\partial
\tilde{p}_{i}}\frac{\partial G}{\partial \tilde{x}^{i}})
\end{eqnarray}
where the constant parameters $\theta_{ij}$ of the
noncommutativity is real and antisymmetric, which has dimension of
area.

In our formulation for noncommutative black hole, one can still
use the usual definition for the metric
\begin{eqnarray}\label{metric}
&&ds^{2}=f(\tilde{r})dt^{2}-\frac{d\tilde{r}^{2}}{f(\tilde{r})}-\tilde{r}^{2}(d\theta^{2}+\sin^{2}\theta
d\phi^{2})
\end{eqnarray}
where $\tilde{r}$ satisfies Eq.(\ref{commutation2}). However, one
should be aware that there is no modified Einstein equation in
this case. In our approach, since the noncommutativity parameter,
if it is non-zero, should be very small compared to the length
scales of the black hole (for example, the horizon of black hole),
one can always treat the noncommutative effects as some
perturbations of the commutative counter-part and therefore one
can use the usual metric. $f(\tilde{r})$ in terms of the
noncommutative coordinates $\tilde{x}_{i}$ is:
\begin{eqnarray}\label{coordinate}
&&f(\tilde{r})=1-\frac{2GM}{\sqrt{\tilde{x}_{i}\tilde{x}^{i}}}\;.
\end{eqnarray}

We note that there is also a new coordinate system
\begin{eqnarray}\label{system}
&&x_{i}=\tilde{x}_{i}+\frac{1}{2}\theta_{ij}\tilde{p}_{j}\;,\nonumber\\
&&p_{j}=\tilde{p}_{j}\;,
\end{eqnarray}
where the new variables satisfy the usual Poisson brackets:
\begin{eqnarray}\label{poisson}
&&\{F,G\}=\frac{\partial F}{\partial x_{i}}\frac{\partial
G}{\partial p^{i}}-\frac{\partial F}{\partial p_{i}}\frac{\partial
G}{\partial x^{i}}
\end{eqnarray}
and
\begin{eqnarray}\label{commutation3}
&&\{x_{i},x_{j}\}=0\;,\nonumber\\
&&\{x_{i},p_{j}\}=\delta_{ij}\;,\\
&&\{p_{i},p_{j}\}=0\;\nonumber.
\end{eqnarray}

Using the new coordinates, we have
\begin{eqnarray}\label{newcoordinates}
&&f(r)=
1-\frac{2GM}{\sqrt{(x_{i}-\theta_{ij}p^{j}/2)(x_{i}-\theta_{ik}p^{k}/2)}}\nonumber\\
&&\qquad\,=1-\frac{2GM}{\sqrt{r^{2}-\frac{(\vec{p}\times\vec{\theta})\cdot
\vec{r}}{4}+\frac{\vec{p}^{2}\cdot\vec{\theta}^{2}-(\vec{p}\cdot\vec{\theta})^{2}}{16}}}\\
&&\qquad\,=1-\frac{2GM}{\sqrt{r^{2}-\frac{\vec{L}\cdot
\vec{\theta}}{4}+\frac{\vec{p}^{2}\cdot\vec{\theta}^{2}-(\vec{p}\cdot\vec{\theta})^{2}}{16}}}\nonumber
\end{eqnarray}
where $\theta_{ij}=\frac{1}{2}\epsilon_{ijk}\theta_{k}$ and
$\vec{L}=\vec{r}\times\vec{p}$. It is worth noting that the terms
of above function are not only the powers in $\theta$ but also the
same powers in momenta. The horizon $\tilde{r}_{h}$ of the
noncommutative metric (\ref{metric}) is
\begin{eqnarray}\label{newcoordinates}
&&\tilde{r}_{h}=[4(GM)^{2}+\frac{\vec{L}\cdot\vec{\theta}}{4}+\frac{\vec{p}\,^{2}\cdot\vec{\theta}\,^{2}-(\vec{p}\cdot\vec{\theta})^{2}}{16}]^{\frac{1}{2}}\;,
\end{eqnarray}
which satisfies
\begin{eqnarray}\label{f}
&&f(\tilde{r}_{h})=0\;.
\end{eqnarray}
Because Schwarzschild external metric is a solution for the
gravitational field surrounding a non-rotating mass,
Eq.(\ref{newcoordinates}) should be reduced to
\begin{eqnarray}\label{newcoordinates1}
&&\tilde{r}_{h}=[4(GM)^{2}+\frac{\vec{p}^{2}\cdot\vec{\theta}^{2}-(\vec{p}\cdot\vec{\theta})^{2}}{16}]^{\frac{1}{2}}\;.
\end{eqnarray}

 The Hawking temperature and the horizon area of
Schwarzschild black hole in noncommutative spaces are respectively
$T_{H}=\frac{GM}{2\pi \tilde{r}_{h}^{2}}$ and
$A=4\pi\tilde{r}_{h}^{2}$. Obviously, they depend on the phase
coordinate $p_{i}$ because of the effect of noncommutative space.
For a rest observer, $p_{i}=0$ so that $\tilde{r_{h}}=2GM$, and
the horizon area increases with the momentum $|\vec{p}|$. In the
limit $\theta_{i}\rightarrow 0$, the horizon is reduced to one in
commutative space.

 \vspace{0.8cm} \noindent\textbf{ Acknowledgments:} This work is
supported by National Natural Science Foundation of China under
Grant No. 10473007.

\end{document}